\documentclass[%
reprint,
 amsmath,amssymb,
 aps,
prstab,
floatfix,
]{revtex4-1}

\usepackage{amsmath}
\usepackage{amssymb}
\usepackage{latexsym}
\usepackage{graphicx}% Include figure files
\usepackage{dcolumn}% Align table columns on decimal point
\usepackage{bm}% bold math
\begin{document}

\title{Direct measurements of air layer profiles under impacting droplets using high-speed color interferometry}
\author{Roeland C. A. van der Veen}
\author{Tuan Tran}
\email{t.tran@utwente.nl}
\author{Detlef Lohse}
\email{d.lohse@utwente.nl}
\author{Chao Sun}
\email{c.sun@utwente.nl}
\affiliation{Physics of Fluids, University of Twente, P.O. Box 217, 7500 AE Enschede, The Netherlands}

\date{\today}% It is always \today, today,

\def\Re{\¤{Re}}
\def\We{\textrm{We}}
\def\bsa{\boldsymbol{a}}
\def\bsb{\boldsymbol{b}}
\def\bsl{\boldsymbol{L}}

%%%%%%%%%%%%%%%%%%%%%%%%%%%%%%%%%
\begin{abstract}
A drop impacting on a solid surface deforms %even 
before the liquid makes contact with the surface.
% due to the pressure 
%increase in the air layer under the drop. 
%This deformation starts at a very short 
%time before the impact and effectively flattens 
%the lower surface of the drop. 
%The result is that
%the drop starts spreading on a very thin layer of air
%before the liquid wets the surface. 
%In this study, 
We directly measure the
time evolution of the air layer profile under the droplet
using high-speed color interferometry,
obtaining the air layer thickness 
%absolute thickness of this air layer using high-speed color interferometry.
%We obtain the evolution of the air thickness
before and during the wetting process. 
Based on the time evolution of the extracted profiles obtained at multiple times, 
we measure the velocity of air exiting from the gap between 
the liquid and the solid, and 
 account for the wetting mechanism and bubble entrapment.
The present work 
offers a tool to accurately measure the air layer profile and  
quantitatively study the impact dynamics at a
short time scale before impact.
\end{abstract}

\maketitle

%\section{Introduction}
%%%%%%%%%%%%%%%%%%%%%%%%%%%%%%%%%

Drop impact on solid 
surfaces, beside its inherent beauty,
has been playing an increasingly important
role in industrial processes as diverse as
ink-jet printing, spray cooling, and spray coating.
%Since it was first studied by \cite{worthington08}, 
%the phenomenon has been studied extensively 
%(see review by \cite{yarin06}), yet 
%is still far from being fully understood and continues
%to chalenge physicists. 
Since it was first studied in 1876 by Worthington \cite{worthington76},
the phenomenon has received tremendous attention from
researchers, yet our understanding of this subject 
is still far from being complete
(see review article \cite{yarin06}).  
A challenge in studying this problem %mostly 
arises from 
widely different time and spatial scales
of the involved effects.
%an array of involving phenomena 
%occurring at widely different 
%time and spatial scales. 
Another difficulty comes from determining relevant 
physical parameters that govern the impact dynamics.
For example, beside apparent parameters such as 
the surface roughness and wettability,
the liquid
viscosity, surface tension, and density, 
it was 
%not until recently that \cite{xu05} 
 recently discovered
that the ambient pressure %also comes into play
%by experimentally showing that it 
is also a crucial parameter as it dictates the splash threshold 
after impact \cite{xu05}. 
This finding and subsequent studies \cite{driscoll11,thoroddsen11}
suggest that the air layer 
between an impinging droplet and a solid surface 
%in determining the 
may have significant effects on the
impact's outcomes.
%It also shows that impact of droplet consists of 
%an array of 
%connecting phenomena among which
%the first one may have crucial effects on the last one.
Hence, it is essential 
to understand how the drop and the surface interact
through the air layer.

On the theoretical side, a mechanism of splash formation
focusing on the short time scale within which 
the drop starts being deformed has been proposed  
\cite{mandre09,mani10}. 
Detailed analysis and simulations have
been subsequently developed \cite{hicks10,brenner11}.
On the experimental side, the dynamics of droplet impact
at the earliest time scale have also been studied;
one of the most remarkable phenomena is the detection 
of entrapped bubbles under an impacting drop 
\cite{chandra91,thoroddsen98,vandam04,thoroddsen05}.
The existence of these bubbles indicates
that the drop's bottom surface is deformed before it 
makes contact with the surface. There is, however, a 
lack of detailed measurements of the air layer thickness
at the earliest time of impact,
%when the drop starts being 
%deformed, 
as well as the 
formation of the entrapped bubbles.

Here we report the first direct measurement of the evolution 
of the air layer profile
 between 
an impinging droplet 
and a solid surface 
using high-speed color interferometry. 
We focus on the earliest time of impact when 
the liquid has not touched the surface 
but starts being deformed
due to the pressure increase 
in the air layer 
between the liquid and the solid surface.
%We provide the first experimental quantification
We measure the air flow between the droplet and the solid surface,
and investigate the mechanism 
of bubble entrapment.
% which was 
%observed in many previous studies. 

%%%%%%%%%%%%%%%%%%%%%%%%%%%%%%%%%%%%%%%%%%%%%%
%%%%%%%%%%%%%%%%%%%%%%%%%%%%%%%%%%%%%%%%%%%%%%
%%%%%%%%%%%%%%%%%%%%%%%%%%%%%%%%%%%%%%%%%%%%%%
%\section{Experiments}
%
\begin{figure}[ht!]
\begin{center}
  \includegraphics[width=7cm]{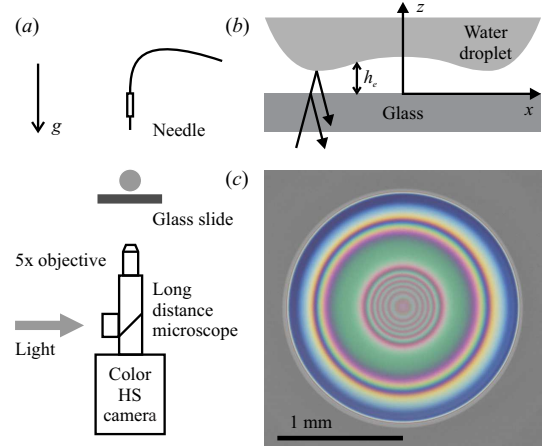}
    \caption{
    ({\it a}) Schematic of the experimental setup (not drawn to scale)
used to study droplet impact on smooth surfaces.
A water droplet of initial diameter $D_0\approx 2\,$mm falls on a glass 
slide of average roughness $10\,$nm.
The bottom view is captured by a high-speed color camera (SA2, Photron Inc.). 
%high-speed cameras respectively. 
The camera is connected to a long working-distance
microscope and a 5x objective to obtain a $2\,$mm field of view. 
 ({\it b}) Schematic of the air film between the drop and the glass 
 slide (not drawn to scale). Light is supplied from 
 the bottom for illumination; 
 reflection of light from 
 the upper surface of the glass slide 
 and from the bottom surface of the drop causes  
interference fringes captured with the color camera. 
 ({\it c}) An example of an interference pattern.}
\label{fig:setup}
   \vspace{-0.5cm}
\end{center}
\end{figure}
In Fig.~\ref{fig:setup}a, 
we show a schematic of the experimental setup for the present work.
% to study droplet impact on smooth surfaces. 
%An experimental setup is sketched in Fig. \ref{fig:setup}a.
We generate 
liquid drops 
by using a syringe pump 
to push liquid out of 
a fine needle.  
The drop detaches 
as soon as its weight overcomes the surface tension 
and then falls on a microscope glass slide 
(Menzel microscope slide, average roughness $\approx$ 10nm). 
In our experiments, 
the working liquid is milli-Q water 
(density $\rho_w = 998\,$kg/m$^3$,
surface tension $\sigma_w = 72\times 10^{-3}\,$N/m,
viscosity $\nu_w = 10^{-6}\,$m$^2$/s).
The drop typically has diameter $D\approx2\,$mm
and its velocity before impacting the surface
can be adjusted by varying the needle's height $H$.
We capture the drop impact
from the bottom
%From the side-view recording 
%of the drop before impacting the surface 
%(obtained by a monochromatic 
%high-speed camera, SA1, Photron Inc.), 
%we measure its diameter $D$ and impact velocity $V$.
%The bottom-view recording of the impact area is obtained 
with a color high-speed camera 
 (SA2, Photron Inc.) connected 
to a long working-distance microscope (Navitar Inc.)
 and a 5X objective. 
 The field of view achieved by this combination is $2\,$mm. 
We illuminate the impact area from below
by supplying white light 
from a high-intensity fibre lamp (Olympus ILP-1) 
to the microscope's coaxial light port. 
When a drop approaches the glass slide, 
a thin film of air is formed between the liquid and solid
surfaces before wetting occurs.  
Light of the same wavelength coming from the bottom,
upon reflection from both surfaces of the film (Fig.~\ref{fig:setup}b)
forms interference patterns recorded by the 
%bottom-view 
camera.
Each one of these patterns consists of constructive (bright)
and destructive (dark) fringes; the fringe spacing 
depends on the air layer thickness and
the wavelength of incident light.
Since the lamp emits light of multiple wavelengths,
the superposition of all available patterns
produces concentric rings of rainbow colors 
as shown in Fig.~\ref{fig:setup}c.
%
%Upon reflection from both surfaces of the film (figure \ref{fig:setup}c), 
%light of the each wavelength forms an interference
%pattern which consists of constructive (bright)
%and destructive (dark) fringes. The spacing 
%of these fringes depends on the air layer thickness and
%the wavelength of incident light.  
%Since the lamp emits light of multiple wavelengths,
%each of which forms a distinct pattern;
%the superposition of all available patterns
%produces concentric rings of rainbow colors 
%as shown in figure \ref{fig:setup}(d).

In most of our experiments,
we set the camera's frame rate to 10000 
frame per second (fps),
and its resolution to $512\times512\,$pixels
to capture droplets with 
impact velocity less than $0.5\,$m/s.
In the case that the impact velocity is higher,
the frame rate can be set as high 
as 86400 fps 
at resolution $32\times256\,$pixels to 
capture the impact dynamics.
 
%%%%%%%%%%%%%%%%%%%%%%%%%%%%%%%%%%%%%%%%%%%%%%
%%%%%%%%%%%%%%%%%%%%%%%%%%%%%%%%%%%%%%%%%%%%%%
%%%%%%%%%%%%%%%%%%%%%%%%%%%%%%%%%%%%%%%%%%%%%%
%\section{Methods}
%\label{sec:Method}
%
% First of all, we explain
% our method to extract the absolute thickness of 
%the air layer between an impinging drop 
%and a glass surface. 
%In principle, the method is based on a ``color matching'' approach
%that consists of two main steps. 
%In the calibration step, we generate 
%a set of reference colors
%from an interference pattern of 
%an air film with known geometry.
%A reference color is then 
%associated with an unique value of thickness.
% The second step is color matching.
%For each point in the pattern
%caused by an impinging droplet, we search
%for its best color match in the database. 
%The film thickness at that point
% can then be determined
% using the color-thickness relation 
% established in the first step.
 %
%%%%%%%%%%%%%%%%%%%%%%%%
%%%%%%%%%%%%%%%%%%%%%%%%
%%%%%%%%%%%%%%%%%%%%%%%%
 %\subsection{Absolute height calibration of reference colors}
 \begin{figure}
  \centerline{\includegraphics[width=7cm]{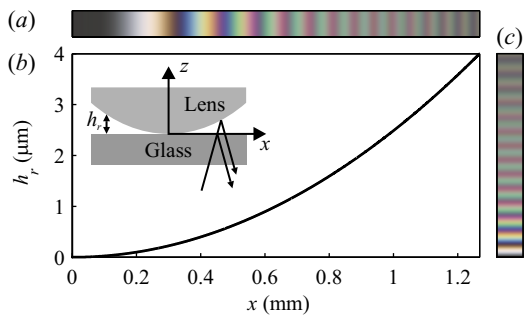}}
  \caption{
    ({\it a}) Reference thickness of the air film between 
    a lens and a glass slide. The lens has radius of the surface
    in adjacent with the glass $R=200\,$mm. 
    Inset: schematic of the setup used to calibrate colors.
     ({\it b}) Color variation in the radial direction of the interference 
     pattern used to calibrate colors.    
    ({\it c}) Relation between thickness and reference colors.
     }
        \vspace{-0.5cm}
\label{fig:calibration}
\end{figure}
 In order to extract the absolute thickness of 
the air layer between an impinging drop 
and a glass surface, we construct a 
set of reference colors 
 that can be related
 to absolute thickness.
 We put a convex lens 
 on top of the glass slide (see inset in Fig.~\ref{fig:calibration}b)
 and observe
 the interference rings caused by 
 the air film between two surfaces.
 Since the pattern consists of concentric 
 rings of different colors, and the air 
 film thickness is known at each radial 
 location, each color along a line 
 passing through the center of these rings 
 is associated with a thickness value.
 In Fig.~\ref{fig:calibration}b, we show the 
air thickness profile between the lens and
the glass slide. The color variation due to change
in air thickness is 
obtained by taking a thin radial strip 
of $100\times2200\,$pixels
from an image of an interference pattern 
and then averaging colors 
in the transverse direction to reduce noise.
The resulting strip (Fig.~\ref{fig:calibration}a),
which has no color variation in the transverse direction,
contains $N=2200\,$pixels in the $x-$direction
and hence $N$ reference colors
 that can be used for calibration.
  Since the camera
 uses the sRGB model
to represent colors,
the color of each pixel $i$ is represented by a color vector $(R_i,G_i,B_i)$.
The pixel's coordinate is $x_i$, which is 
  related to a value of thickness $h_r^i$.
  Thus, we have a set of reference colors $(R_i,G_i,B_i)$
  for $1\le i \le N$, each of which is associated with 
  a reference thickness $h_r^i$. The
reference thickness range is $0\le h_r^i \le 4\,\mu$m. 
  The thickness-color relation is shown
 in Fig.~\ref{fig:calibration}c.

%\section{Results and discussion}
%
\begin{figure}
  \centerline{\includegraphics{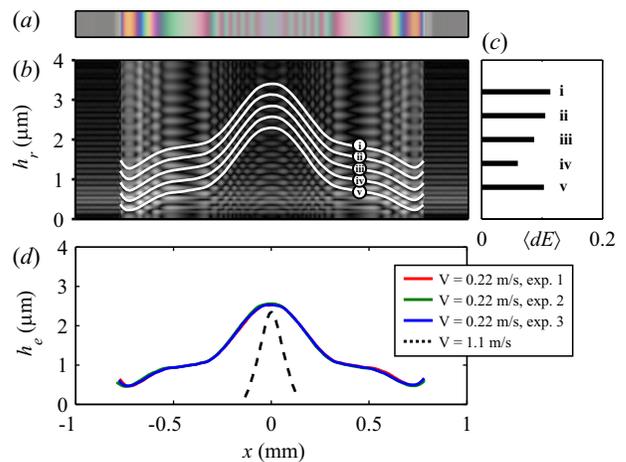}}% Images in 100% size
  \caption{
   ({\it a}) Color sample of interference pattern taken at $t=1.67\,$ms after
   the bottom-view camera detected the drop. 
    ({\it b}) Color difference in grayscale computed by Eq. \ref{eqn:color_dif}
   with candidate profiles shown in white solid lines.
   ({\it c}) Average color difference of candidate profiles shown in ({\it b}).
   ({\it d}) Solid lines: profiles computed from three different experiments with $V=0.22\,$m/s.
   Dashed line: $V=1.1\,$m/s. Note the extremely different length scales at the $x-$axis (mm) and $h_r-$axis
   ($\mu$m) in ({\it b}) and ({\it d}).
     }
        \vspace{-0.5cm}
\label{fig:result1}
\end{figure}
\begin{figure*}[ht!]
  \centerline{\includegraphics[width=14cm]{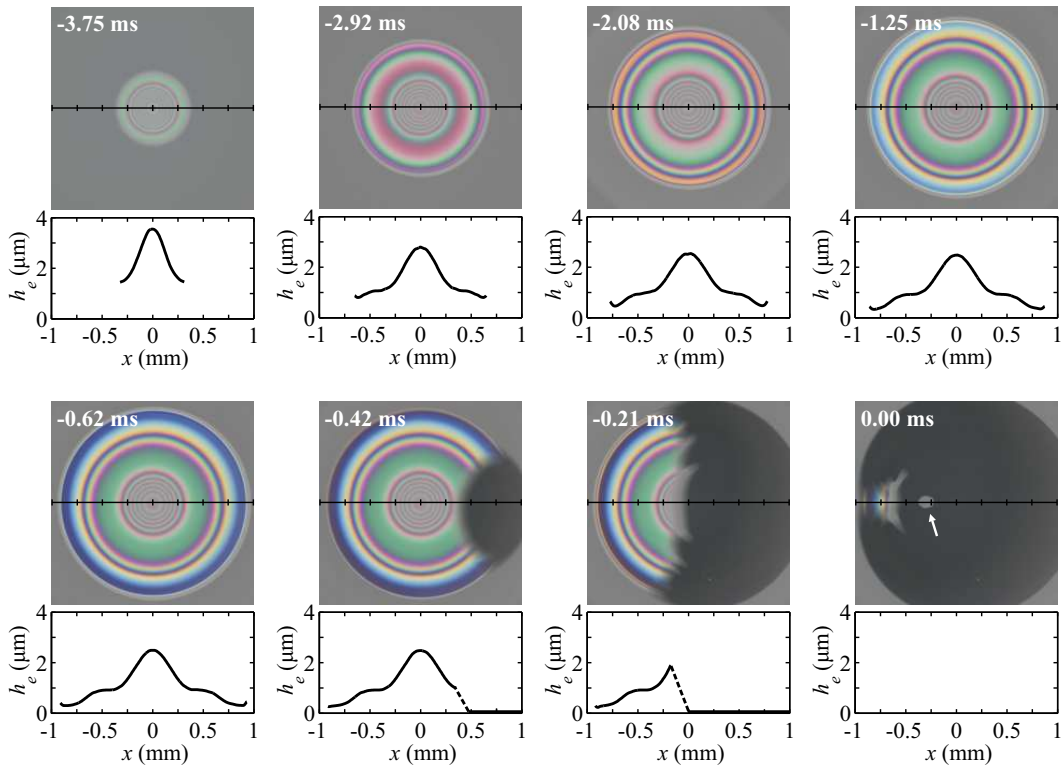}}% Images in 100% size
  \caption{
  Snapshots of interference patterns obtained during 
  drop impact and their corresponding calculated profiles 
  ($V=0.22\,$m/s, $D=2\,$mm). See the supplementary movie.
  }
     \vspace{-0.5cm}
\label{fig:series}
\end{figure*}
\begin{figure}[h!]\centerline{\includegraphics[width=7cm]{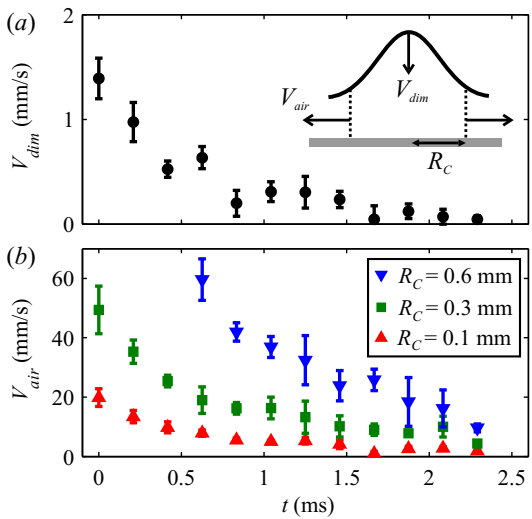}}% Images in 100% size
  \caption{
  ({\it a}) Dimple velocity $V_{dim}$ vs. time. 
  ({\it b})  
  The average velocity of air $V_{air}$ vs. time
    at different radial locations $R_C=0.1\,$mm (upward triangles), $R_C=0.3\,$mm (squares), $R_C= 0.6\,$mm (downward tringles) for an experiment with impact velocity $V=0.22\,$m/s. 
     }
   \vspace{-0.5cm}
\label{fig:dimple}
\end{figure}
The sRGB model, however, is generally not preferred when
comparing colors between experiments
because it does not decouple light
intensity and color information,
which poses a problem due to variations in illumination conditions such as 
light intensity, incident and observing angles.
Instead, we use the CIE 1976 color model (also called CIELAB),
a 
%{\it device--independent} 
model that is
 most effective in decoupling light intensity (\cite{hunt98})
(see the Supplemental material for details).
 A color in CIELAB model 
has three components: 
$\bsl$ for lightness information,  
and $\bsa$ and $\bsb$ for color information.
Thus, we can separate light intensity 
from our analysis by
omitting the component $\bsl$.
Each reference color $i$ after intensity decoupling 
is represented by a two-component vector $(\bsa_r^i,\bsb_r^i)$
and is associated with a value of reference thickness
$h_r^i$ for $1 \le i \le N$ and 
$h_r^i$ is in the range $0\,\mu{\rm m} \le h_r^i \le 4\,\mu{\rm m}$.  

In Fig.~\ref{fig:result1}a, we show a color sample, which was taken along
a diameter of 
an interference pattern under a drop with impacting velocity
$V=0.22\,$m/s. 
%We demonstrate this process for a representative interference pattern
%taken at $t=1.67\,$ms
%(here we set the time origin $t=0\,$ms when the camera first detects the drop). 
%The color sample %used to extract the air thickness 
%was taken along a diameter of the pattern 
%%and is
After converting the color of each pixel to CIELAB color space and
decoupling light intensity, we calculated the color difference $dE^{ij}$
for each color $(\bsa_e^j,\bsb_e^j)$ in the sample
($1\le j \le M=596$)
and each reference color $(\bsa_r^i,\bsb_r^i)$
using the Euclidean distance:
 \begin{equation}
dE^{ij} = \left [ (\bsa_e^j - \bsa_r^i)^2 + (\bsb_e^j - \bsb_r^i)^2 \right ]^{1/2}, 
\label{eqn:color_dif}
 \end{equation}
for $1\le i \le N$ and $1\le j \le M$.
Since each color $(\bsa_e^j,\bsb_e^j)$ is associated 
 with a coordinate $x_j$ and, recall that each 
 reference color $(\bsa_r^i,\bsb_r^i)$ is associated with a
value of reference thickness $h_r^i$, the color difference 
$dE^{ij}$ can be thought of as a function of $h_r^i$ and $x_j$.
In figure \ref{fig:result1}b, we show 
a plot of $dE^{ij}$ in grayscale for 
$1 \le i \le N$ and for $1\le j \le M$.
The range of the index 
$i$ translates to
the range of reference thickness 
as $0 \le h_r \le 4\,\mu{\rm m}$,
and $j$ to $-1 \le x \le1\,$mm.
In the plot, black means $dE=0$ and hence 
zero color difference, whereas white means
the largest color difference.
A vertical line at a particular value of $x$ 
has all possible values of 
the film thickness at that point; the correct
thickness value corresponds to the darkest 
point. 
In the case that there are multiple dark points 
on the same vertical line with insignificant difference 
between them, thickness determination
is not trivial. 
We note that, however, the film profile is
continuous and smooth. 
Evidently, there are only a few continuous
dark lines that can be distinguished 
without any abrupt change in slope.
In figure \ref{fig:result1}b, we show the candidate
profiles in white solid lines (labeled
from ({\romannumeral 1}) to ({\romannumeral 5})).
The film thickness profile can be identified by 
considering the average color difference $\langle dE\rangle$
along each candidate profile $L$:
 \begin{equation}
\langle dE\rangle^L = \frac{1}{N_L}\displaystyle\sum_{L} dE^{L},
 \end{equation}
where the sum is taken for all the pixels along the profile $L$ and then
divide by the number of pixels ($N_L$).
 In figure \ref{fig:result1}c, we show $\langle dE \rangle^L$
 for all profiles. 
 The smallest color difference is along profile 
$({\romannumeral 4})$ for which $\langle dE\rangle^{\romannumeral 4}=0.06$, whereas the 
second smallest one is along profile $({\romannumeral 3})$
for which $\langle dE\rangle^{\romannumeral 3}=0.09$.
As a result, we conclude that 
profile $({\romannumeral 4})$ is the air layer profile.  
A test case of an air film with a known thickness profile shows that
the accuracy of our method is within $40\,$nm (see the Supplemental material).
To check the reproducibility
we repeated the experiment several times and 
extracted the air thickness in each experiment
at the same time. The computed profiles are shown
in Figs.~\ref{fig:result1}d. Given the variations between experiments
such as releasing time, drop size, surface properties, etc.,
the method gives remarkably consistent results.

In figure \ref{fig:series}, we show 
interference patterns obtained during 
drop impact ($V=0.22\,$m/s and $D=2\,$mm) 
and their corresponding 
thickness profiles of the air layer.
%The first pattern (at $t=0\,$ms) is the first one detected by the camera
We define $t=0$ as the moment when the liquid completely wets 
the solid surface.
From the first pattern detected by the camera ($t=-3.75\,$ms), 
it is readily seen that a dimple 
%approximately
%$4\,\mu$m high and $0.5\,$mm in the radial 
%direction is formed. 
is already formed, which means that the 
camera did not capture the entire 
deformation process of the drop's lower surface, 
probably due to limited coherence length of the 
light source used in the present experiment. 
Subsequent profiles show
that the dimple's height gradually reduces, 
while the liquid continues 
spreading in the radial direction. 
At $t=-0.42\,$ms, the liquid starts wetting the
glass surface at one point along the rim of the air layer and
then propagates to the other side; 
the wetting process happens faster 
along the rim where the air thickness  is smallest 
and finally traps air bubbles
(indicated by an arrow at $t=0\,$ms). 
For experiments done under the same conditions, 
although the time that wetting occurs 
varies, we observe the same 
bubble-trapping dynamics, that is, 
the wetting front propagates faster at the rim
and finally enclose the air pocket 
underneath the drop. 

%\begin{figure}[h!]
%  \centerline{\includegraphics[width=8cm]{../Figs_prl/Figure8.eps}}% Images in 100% size
%  \caption{
%Thickness profiles of the air layer under drops impacting the surface at
%$V=0.22\,$m/s and $V=1.1\,$m/s. The profiles were calculated from 
%interference snapshots just before wetting starts.}
%\label{fig:2vel}
%\end{figure}

We note that there is a plateau (at $x\approx \pm 0.5\,$mm)
in the thickness profiles from $t=-2.92\,$ms to $t=-0.21\,$ms.
This is due to the drop's oscillation  
as it falls down at a small distance from the surface (in this case
$H=6\,$mm). When the drop detaches from the needle, capillary waves 
are generated and propagate to the other side. The surface deformation 
caused by these waves affects the dimple shape 
in addition to the pressure increase in the air layer under the drop.
In the case that the drop is released from a larger height leaving 
sufficient time for viscosity to damp capillary waves, 
we do not observe the plateau in the thickness profiles.
As shown in Fig.~\ref{fig:result1}d, the plateau is not 
present in the case of higher impact velocity case ($V=1.1\,$m/s,
$H=66\,$mm).

% we show the thickness profiles  
%resulting from experiments with different  
%impact velocity, both measured before wetting occurs: $V=0.22\,$m/s
%(releasing height $H=6\,$mm) and 
%
%Indeed, the plateau is only present in the case of small releasing height.

%An estimate of the time scale associated with the capillary wave
%($\tau\sim\sqrt{\rho R^3/\sigma}\approx3\,$ms) indeed 
%shows that it is comparable 
%to the impact time measured in our experiment 
%(approximately $3\,$ms).

We now quantify the velocity at the center of the dimple $V_{dim}$ 
(inset of Figs.~\ref{fig:dimple}a).
As shown in Figs.~\ref{fig:dimple}a, 
 $V_{dim}$ is found to be very small (roughly two orders of magnitude smaller)
as compared to the impact velocity $V=0.22\,$m/s, which 
implies that the fluid at the bottom of the drop has decelerated 
before the camera starts capturing the interference fringes.
Nonetheless,
our measurements capture well the 
 deceleration process of the lower surface of the drop from the detection point until
 it is brought to rest.
Moreover, we estimate the horizontal velocity of air $V_{air}$
based on the change in volume confined by a cylinder of radius $R_C$
under the liquid surface
(inset of Figs.~\ref{fig:dimple}a).
 In figure \ref{fig:dimple}b, we show $V_{air}$ at several 
 values of $R_C$. 
 The data show a consistent increase of the air velocity at a given time
as it gets closer to the rim of the air layer where the thickness 
is minimum.
For higher impact velocity cases, the velocity of air is much higher 
due to the extremely thin air gap at the rim.
%We, however, do not observe an exceeding velocity of the air 
%compared to the impact velocity $V$. This is probably due to the 
%small $V$ used in our experiments. 
%For an extended range of 
%impact velocity, a systematic study of the air velocity is to be investigated.

%
In conclusion, we have used high-speed color interferometry to 
measure the complete profile and its evolution 
of the air layer
under an impacting drop 
for impact velocity $V=0.22\,$m/s and $V=1.1\,$m/s. 
%The minimum thickness that can be measured 
%with this method is $200\,$nm, with accuracy within $50\,$nm.  
From the experimental measurements, we account for the wetting mechanism 
which results in entrapment of bubbles after impact. 
We also experimentally quantify the velocity of air flow between the drop and the surface,
as well as the velocity of the dimple before wetting occurs.
Our results offer a benchmark for theories of drop impact.

%High velocity: systematic study is underway
%Compared to the existing studies 

%%%%%%%%%%%%%%%%%%%%%%%%%%%%%%%%%%%%%%%%%%%%%%
This study was financially supported by the 
European Research Council ERC.
%%%%%%%%%%%%%%%%%%%%%%%%%%%%%%%%%%%%%%%%%%%%%%

\vspace{-0.3cm}

%\bibliographystyle{prsty}

%\bibliography{air_thickness}
%merlin.mbs 2010-03-15 4.21a (PWD, AO, DPC)

\end{document}